# EFFICIENT KEY GENERATION FOR DYNAMIC BLOM'S SCHEME

Divya Harika Nagabhyrava

**Abstract**: The need for secure communication between the nodes in a network has led to the development of various key distribution schemes. Blom's scheme is a prominent key exchange protocol used for sensor and wireless networks, but its shortcomings include large computation overhead and memory cost. In this paper, we propose an efficient key management scheme by generating new public and private keys. We also focus on making Blom's scheme dynamic by randomly changing the secret key that is generated by the base station and we propose the use of the mesh array for matrix multiplication for reducing the computation overhead.

**INTRODUCTION**
Recent improvements in the field of electronic and computer technologies, have paved the way for the proliferation of wireless sensor networks (WSN).These sensors are used to collect environmental information and they have been considered for various purposes, including security monitoring, target tracking and research activities in hazardous environments. Encryption, authentication, confidentiality and key management are the main aspects to focus on to achieve security in wireless sensor networks. Different key management schemes have been developed for authentication and confidentiality purposes. Symmetric algorithm base key management protocols are used in wireless sensor networks because key management protocols based on public key cryptography are inefficient due to resource limitations.

Blom's scheme [1] and others related to it [2]-[10] to it are often used for secure sensor communication. Issues that need to be considered include key management [13]-[17] and randomness of sequences [18]-[23]. In this paper we review general considerations and then propose our scheme for dynamic change of keys. We present results of our simulations.

**GENERAL CONSIDERATIONS**
The main challenges of the sensor nodes are as follows:

- ➢ The nodes are solely responsible for any change in the configuration as there would be no human intervention once the sensor network is deployed.
- ➢ There is only a finite source of energy as the nodes are not connected to any other energy source. Hence, communication should be minimized.
- ➢ It is required that a sensor network system be adaptable to changing connectivity as well as changing environmental stimuli.

Sensor networks have various requirements:

*Integrity*: Data integrity ensures that the data received by the recipient node is same as the data sent by the sender node and is not altered or modified by any adversary. Different cryptography techniques can be employed to ensure integrity of data.

*Authentication and Confidentiality*: The two nodes (sender and receiver) must be authenticated before data transmission. It means that the receiving node must be ensured that the data is sent from a trusted node and not from any malicious node. Also, the data that is sent must be received only by the authenticated receiver and must not be exposed to any other nodes.

*Availability, reliability and resiliency:* It ensures proper connectivity of the nodes and that the data and information is available to access at all times whenever and wherever required by the authorized nodes. This ensures protection from attacks such as denial of service attacks, etc. It also ensures that the data packet is delivered to its destination.

*Data freshness*: It ensures that only recent data must be delivered. It can be helpful to detect the adversary nodes as they can send the old messages.

In order to perform better in terms of accuracy, reliability and security there are some limitations that need to be overcome. Some of the limitations are: (1) *Energy:* Sensor network consists of tiny sensors which run on batteries. These batteries have limited power supply. Hence there comes the need to conserve energy and transmit data efficiently without consuming much energy. (2) *Computation:* Due to energy constraints, sensors is also limited by low computational capacity. Hence, algorithms which require large computations must be avoided to incorporate in sensor networks. (3) *Communication:* Sensors are linked via wireless connection, and therefore the bandwidth is often limited. This also limits the transmission of data and information.

**Generating Symmetric Matrix**
In the Blom's scheme, the Central Authority or the base station randomly generates the secret key which is a symmetric matrix. Here, we obtain the secret symmetric matrix by multiplying a random matrix with its transpose.

The secret symmetric matrix may be updated in a variety of ways:

- The symmetric matrix multiplied by its transpose.
- The symmetric matrix added with another symmetric matrix.

which will again be a symmetric matrix that can be used as a secret matrix.

**Mesh Array for Matrix Multiplication**

The mesh array of matrix multiplication [11],[12] was introduced in 1988 to speed up the computation of multiplying two n×n matrices using distributed computing nodes. In contrast to the standard array that requires 3n-2 steps to complete its computation, the mesh array requires only 2n-1 steps. The speedup of the mesh array is a consequence of the fact that no zeros are padded in its inputs. In the modified Blom's scheme, the mesh array for matrix multiplication can be used by the central authority or the base station for generating the private matrix and by the user pair in order to obtain the shared key by multiplying the public and private key.

Consider two 4*4 matrices:

$$A=\begin{bmatrix} a11 & a12 & a13 & a14 \\ a21 & a22 & a23 & a24 \\ a31 & a32 & a33 & a34 \\ a41 & a42 & a43 & a44 \end{bmatrix} \quad B=\begin{bmatrix} b11 & b12 & b13 & b14 \\ b21 & b22 & b23 & b24 \\ b31 & b32 & b33 & b34 \\ b41 & b42 & b43 & b44 \end{bmatrix}$$

The mesh architecture for multiplying the above two matrices (C=AB) is given by:

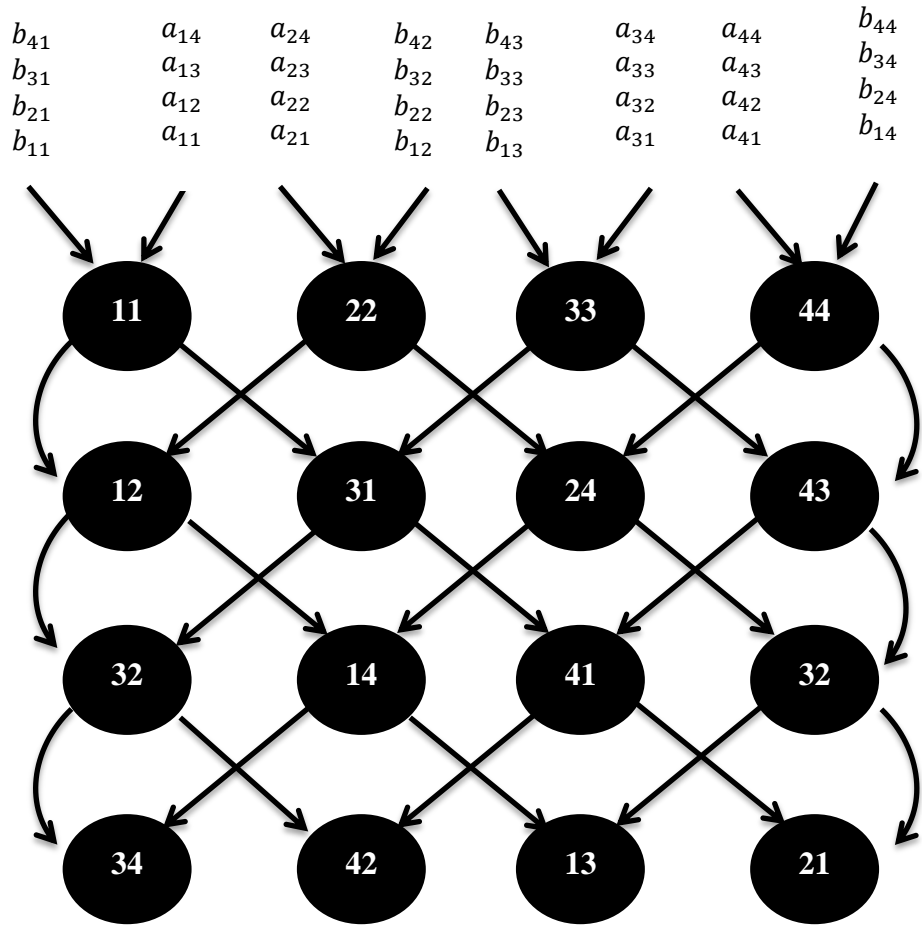

Figure-1: Mesh Array for Matrix Multiplication

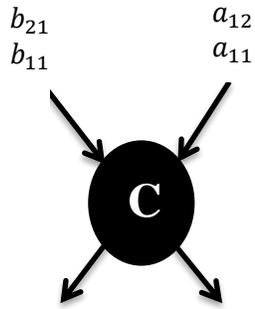

Figure-2: Computing node that generates c= $a_{11}b_{11}+a_{12}b_{21}$

Properties of Mesh Array for Matrix Multiplication:

- For a matrix with odd order, the rows 2 to (n+1) /2 are mirror image of rows (n+3) /2 to n.

- For a matrix with even order, the rows 2 to n/2 are mirror reversed image to rows n/2+2 to n, and the middle row (n/2+1) has self-symmetry.

**BLOM'S SCHEME**

Blom's scheme [1] is a symmetric threshold key exchange cryptography protocol. It allows any pair of users in the system to find a unique shared key for secure communication.

- In this scheme, a network with N users and a collusion of less than t+1 users cannot reveal the keys which are held by other users. Thus, the security of the network depends on the chosen value of t, which is called Blom's secure parameter (t<<N).
- Larger value of t leads to greater resilience but a very high t value increases the amount of memory required to store key information.
- *Generation of Public matrix:*

    Initially, a central authority or base station first constructs a (t + 1) ×N matrix P over a finite field GF (q), where N is the size of the network and q is the prime number. P is known to all users and it can be constructed using a Vandermonde matrix. It can be shown that any t+1 columns of P are linearly independent when $n_i$, i=1, 2,…N are all distinct.

$$P = \begin{bmatrix} 1 & 1 & 1 & ... & 1 \\ n_1 & n_2 & n_3 & ... & n_N \\ n_1^2 & n_2^2 & n_3^2 & ... & n_N^2 \\ ...... & ...... & ...... & ...... & ...... \\ n_1^t & n_2^t & n_3^t & ... & n_N^t \end{bmatrix}$$

- *Generation of Secret Key (Private matrix):*

The central authority or the base station selects a random $(t + 1) \times (t + 1)$ symmetric matrix S over GF (q), where S is secret and only known by the central authority.

A $N \times (t + 1)$ matrix $A = (S.P)^T$ is computed which is needed for generating the shared key.

$$K = A.P = (S.P)^T.P = P^T.S^T.P = P^T.S.P = (A.P)^T = K^T$$

- *Generation of shared key by the user pair:*

    User pair (i, j) will use $K_{ij}$, the element in row i and column j in K, as the shared key. Because $K_{ij}$ is calculated by the i-th row of A and the j-th column of P, the central authority assigns the i-th row of A matrix and the i-th column of P matrix to each user i, for 1, 2….. N. Therefore, when user i and user j need to establish a shared key between them, they first exchange their columns of P, and then they can compute $K_{ij}$ and $K_{ji}$, respectively, using their private rows of A. It has been proved in [10] that the above scheme is t-secure if any $t + 1$ columns of G are linearly independent.

- The t-secure parameter guarantees that no compromise of up to t nodes has any information about $K_{ij}$ or $K_{ji}$.

$A = (S.P)^T \qquad\qquad P \qquad\qquad (S.P)^T P$

$$\begin{bmatrix} i_1 & i_2 & . & . & i_{t+1} \\ & & & & \\ j_1 & j_2 & . & . & j_{t+1} \end{bmatrix} \begin{bmatrix} i_1 & j_1 \\ i_2 & j_2 \\ . & . \\ . & . \\ i_{t+1} & j_{t+1} \end{bmatrix} \begin{bmatrix} & K_{ij} \\ & \\ K_{ji} & \end{bmatrix}$$

Figure-3: Generating keys in Blom's Scheme

**Proposed Algorithm for Distributing Keys:**

In the original Blom's scheme all the computations that are involved in generating the keys are based on a Vandermonde matrix [24] which is a public matrix (P) and known to even the adversaries. The security of the network depends on the secure parameter t. However, for large values of t, number of rows in the matrix increases and which in turn corresponds to a greater value in the columns because the column values increase in a geometric series.

The proposed method makes use of the original Blom's scheme. In the Blom's scheme for any two nodes to generate a shared key, each node should store its private key and the public key of the other node. Since every sensor node is provided with limited memory and energy, it will be difficult to store both the row and column in the sensor memory for a large network. So, the goal is to reduce the memory and computation overhead. To achieve this in Blom's scheme, instead of using Vandermonde matrix we propose the use any random matrix as the public matrix. The overall computation cost can also be reduced because the cost of generating a random matrix is less as compared to the cost of generating a Vandermonde matrix. Also, a random prime number q is chosen. Now, similar to Blom's scheme the operation which are to be performed to generate the keys will depend on the prime number, i.e., the number which depends on the desired key length.

Following are the steps involved in calculating the shared key using the proposed scheme:

- *Generation of Public matrix:*
  Initially, any random matrix of order $(t+1) * (t+1)$ is chosen as the public matrix.
- *Generation of Secret matrix (symmetric matrix):*
  The central authority calculates a $(t + 1) \times (t + 1)$ symmetric matrix S, where S is secret and only known by the central authority.
- *Generation of Private Matrix for obtaining the shared key:*
  A matrix of order $N \times (t + 1)$, $A = (S. P)^T$ is computed which is used for generating the shared key. The base station stores each row of the matrix A in the node memory with corresponding index. This is shown in Figure 4.

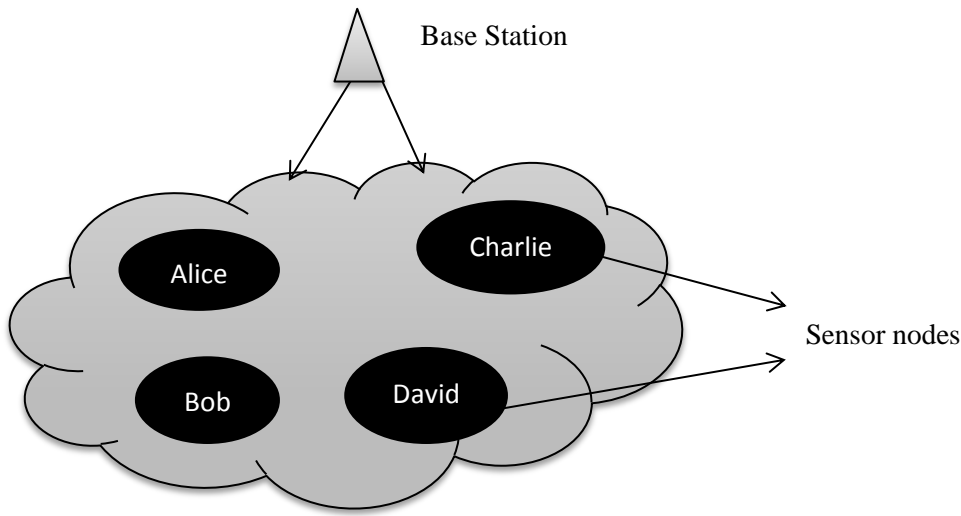

Figure-4. The network

- *Assigning Unique ID:*
  The Central Authority assigns unique id's to all the nodes in the network, which is public to the entire network before the public and private matrices are calculated. Once the node receives its unique id, it responds to the CA with an acknowledge type message or a reply.
  A question here arises, what if there is already an intruder in the network before assigning id's or how to distinguish a new node that enters a network as a trusted node or a malicious node. For this reason, we assume that every node is authenticated in some other network before it enters this network. The following handshake takes place before the node enters any network.

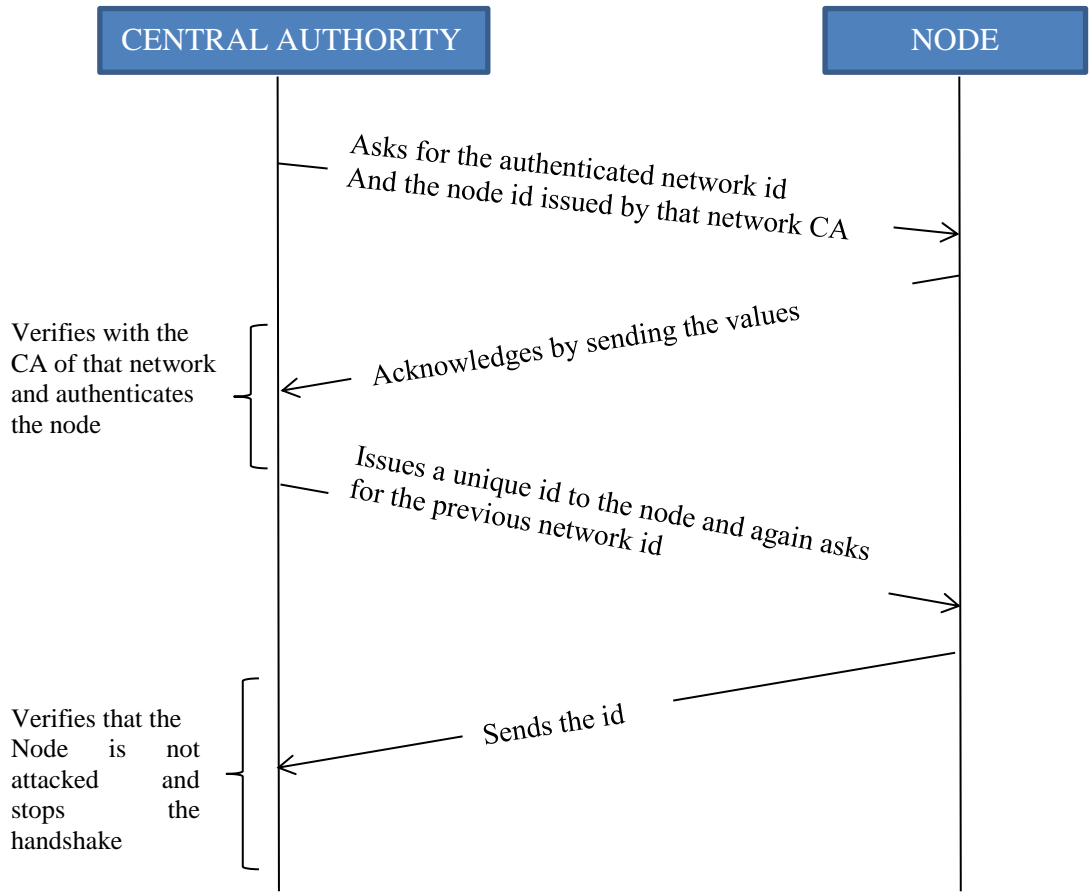

Figure-5 New Node Authentication

- Now, if suppose Alice and Bob want to communicate with each other, they require the private key from the central authority to calculate the shared key. In order to obtain the key, the following handshake takes place.

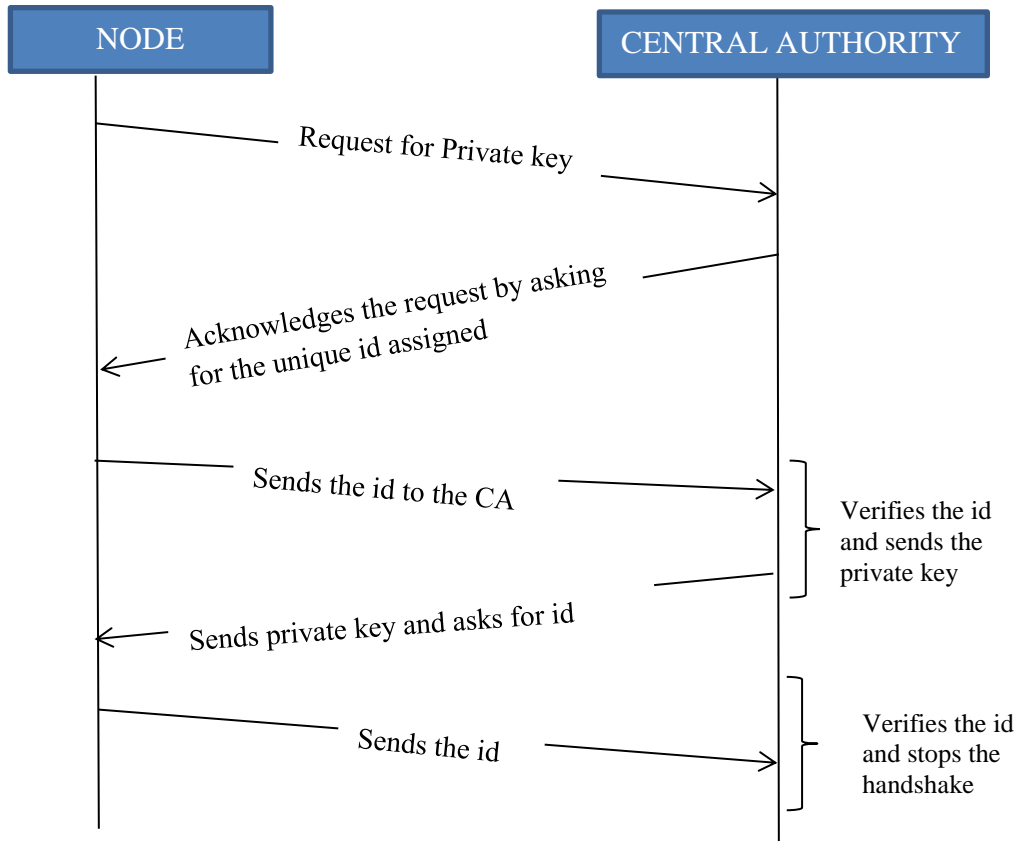

Figure-6 Private key Distribution

- The CA maintains the following intrusion table for the malicious nodes.

| ID (malicious nodes) | INTRUSION COUNT |
|---|---|
| - MN1<br>- MN2 | - 1<br>- 2 |

TABLE-1 Intrusion Table

- In any of the above steps, if the Central Authority detects any node to be malicious, it then updates the unique id value of that node as MN1, increments the intrusion count by 1, blocks all the communication in that particular path and multicasts it to all the nodes in the network.
- *Generation of shared key:*
  Finally user pair (Alice, Bob) can compute the shared key by multiplying the secret row of matrix A stored in the node with column of the public matrix corresponding to the node index with which it want to communicate. The key generation between any two nodes is shown in the following Figure.

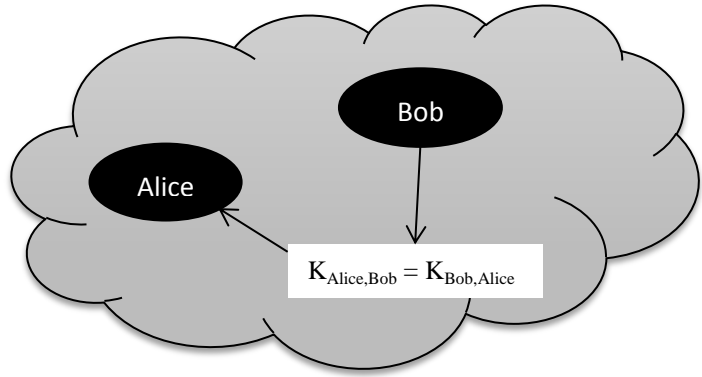

Figure-7. Generating keys

- *Dynamic Secret Matrix:*
    The CA dynamically updates the secret matrix in each of the following cases:
    - ➤ For every increment of the intrusion count in the intrusion table.
    - ➤ Suppose, if node i wants to communicate with node j (both i and j being trusted parties and not intruders). Once, the CA gives the private key to both the nodes, it updates the secret matrix for the next request from the nodes.

The CA does the following handshake at regular time intervals in order to detect the malicious nodes.

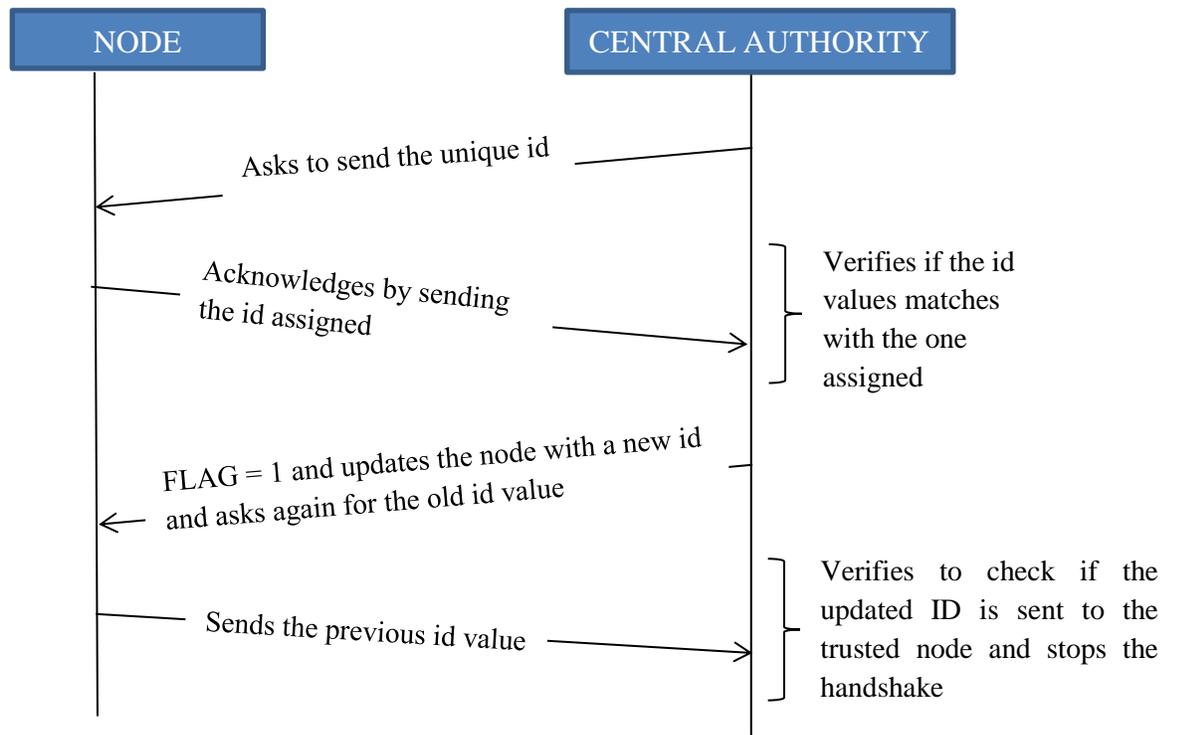

Figure-8: Malicious Nodes Detection

- Once the shared key has been established between the nodes, before the nodes start communicating with each other, the following handshake takes place between the nodes and the Central Authority in order to authenticate each other.
  Let us consider that Alice wants to communicate with Bob:

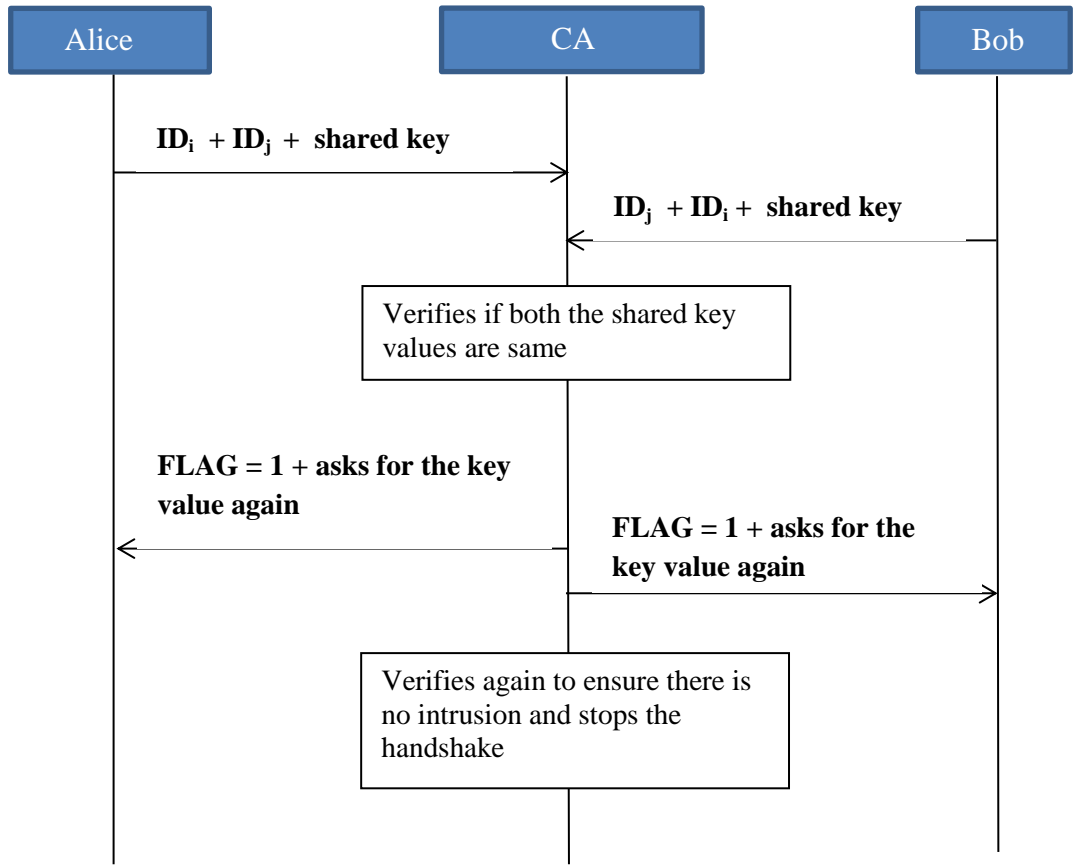

Figure-9: Node Authentication

Further strengthening of the random sequences used can employ the methods of [25],[26].

**Implementation**

The example below shows the working of proposed Scheme that uses a random matrix as public key and generates a private key. Let us consider a network with 4 nodes and the following parameters:

- Let C be the Central Authority or Base Station.
- Secure parameter t = 3, which says if more than 3 nodes in the network are compromised, it is not possible to find the keys of other users.
- Prime number q= 31.
- As the public matrix (P) should be of the order (t+1) * (t+1), P can be taken as any random (3+1) * (3+1) matrix.

$$\begin{bmatrix} 1 & 2 & 3 & 4 \\ 1 & 0 & 1 & 1 \\ 2 & 1 & 3 & 1 \\ 4 & 0 & 9 & 5 \end{bmatrix}$$

- The secret symmetric matrix can be obtained as follows:
  Let us take a random matrix,

$$A = \begin{bmatrix} 1 & 0 & 1 & 1 \\ 1 & 2 & 0 & 1 \\ 0 & 0 & 1 & 1 \\ 0 & 2 & 3 & 1 \end{bmatrix} \quad B = \begin{bmatrix} 1 & 1 & 0 & 0 \\ 0 & 2 & 0 & 2 \\ 1 & 0 & 1 & 3 \\ 1 & 1 & 1 & 1 \end{bmatrix}$$

We obtain the secret matrix S by taking the product of two matrices A and B,

$$S = \begin{bmatrix} 3 & 2 & 2 & 4 \\ 2 & 6 & 1 & 5 \\ 2 & 1 & 2 & 4 \\ 4 & 5 & 4 & 14 \end{bmatrix}$$

- Now, we calculate matrix A using,
  $A = (S.P)^T \mod q$

$$(S.P) = \begin{bmatrix} 25 & 8 & 53 & 36 \\ 30 & 5 & 60 & 40 \\ 23 & 6 & 49 & 31 \\ 73 & 12 & 155 & 95 \end{bmatrix}$$

$$A = (S.P)^T \mod 31 = \begin{bmatrix} 25 & 8 & 22 & 5 \\ 30 & 5 & 29 & 9 \\ 23 & 6 & 18 & 0 \\ 11 & 12 & 0 & 2 \end{bmatrix}$$

Once A is calculated, each sensor node memory is filled with unique row chosen from A with corresponding index. These are the private keys for the nodes.

- Now, for the key generation, we need the public and private keys of the nodes. Suppose, Bob and Charlie wants to communicate with each other. In order to generate the shared secret key, Bob should multiply the private key given by the Central Authority from A with the public column of Charlie in P.

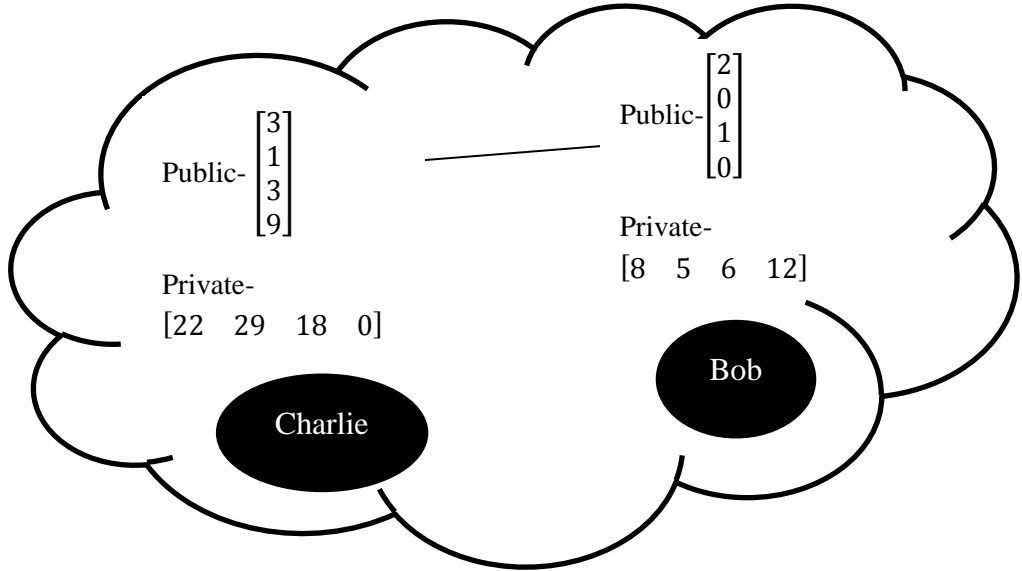

Figure-10: Private and public keys

$$K_{Bob,\ Charlie} = A_{Bob} \cdot P_{Charlie} = [8 \quad 5 \quad 6 \quad 12] \begin{bmatrix} 3 \\ 1 \\ 3 \\ 9 \end{bmatrix} = 155 \bmod 31 = 0$$

$$K_{Charlie,\ Bob} = A_{Charlie} \cdot P_{Bob} = [22 \quad 29 \quad 18 \quad 0] \begin{bmatrix} 2 \\ 0 \\ 1 \\ 0 \end{bmatrix} = 62 \bmod 31 = 0$$

As seen above, both the nodes generate a common key and further communication can be done using the shared key generated.

Let us assume the following:

Alice – 1; Bob – 2; Charlie – 3; David – 4

The table below shows the Public and Private keys for different nodes:

| NODE PAIR | (PUBLIC KEY)$^T$ | PRIVATE KEY | SHARED KEY |
|---|---|---|---|
| $K_{1,2}$ | [2 0 1 0] | [25 30 23 11] | 11 |
| $K_{2,1}$ | [1 1 2 4] | [8 5 6 12] | 11 |
| $K_{1,3}$ | [3 1 3 9] | [25 30 23 11] | 25 |
| $K_{3,1}$ | [1 1 2 4] | [22 29 18 0] | 25 |
| $K_{1,4}$ | [4 1 1 5] | [25 30 23 1] | 22 |
| $K_{4,1}$ | [1 1 2 4] | [5 9 0 2] | 22 |
| $K_{2,3}$ | [3 1 3 9] | [8 5 6 12] | 0 |
| $K_{3,2}$ | [2 0 1 0] | [22 29 18 0] | 0 |
| $K_{2,4}$ | [4 1 1 5] | [8 5 6 12] | 10 |
| $K_{4,2}$ | [2 0 1 0] | [5 9 0 2] | 10 |
| $K_{3,4}$ | [4 1 1 5] | [22 29 18 0] | 11 |
| $K_{4,3}$ | [3 1 3 9] | [5 9 0 2] | 11 |

TABLE-2: Key Generation for all nodes

Now, the CA can anytime update the secret matrix S, to protect the network from malicious users.

$$S' = \begin{bmatrix} 3 & 2 & 2 & 4 \\ 2 & 6 & 1 & 5 \\ 2 & 1 & 2 & 4 \\ 4 & 5 & 4 & 14 \end{bmatrix} \begin{bmatrix} 4 & 5 & 4 & 14 \\ 2 & 1 & 2 & 4 \\ 2 & 6 & 1 & 5 \\ 3 & 2 & 2 & 4 \end{bmatrix}$$

$$S' = \begin{bmatrix} 36 & 37 & 26 & 76 \\ 37 & 32 & 31 & 77 \\ 26 & 31 & 20 & 58 \\ 76 & 77 & 58 & 152 \end{bmatrix}$$

The Private matrix A can be calculated as,

A= (S.P)$^T$ mod q

$$A = \begin{bmatrix} 429 & 439 & 329 & 877 \\ 98 & 105 & 72 & 210 \\ 907 & 929 & 691 & 1847 \\ 587 & 596 & 445 & 1199 \end{bmatrix} \text{Mod } 31$$

$$A = \begin{bmatrix} 26 & 5 & 19 & 9 \\ 5 & 12 & 10 & 24 \\ 8 & 30 & 9 & 18 \\ 29 & 7 & 11 & 21 \end{bmatrix}$$

Now, suppose Bob and Charlie wants to communicate with each other.

$K_{Bob, Charlie} = A_{Bob} \cdot P_{Charlie}$

$$= \begin{bmatrix} 5 & 12 & 10 & 24 \end{bmatrix} \begin{bmatrix} 3 \\ 1 \\ 3 \\ 9 \end{bmatrix}$$

$= 273 \mod 31 = 25$

$K_{Charlie, Bob} = A_{Charlie} \cdot P_{Bob}$

$$= \begin{bmatrix} 8 & 30 & 9 & 18 \end{bmatrix} \begin{bmatrix} 2 \\ 0 \\ 1 \\ 0 \end{bmatrix}$$

$= 25 \mod 31 = 25$

Both the nodes generate a common key and further communication can be done using the shared key generated.

**ANALYSIS AND SIMULATION**

The main drawbacks of Blom's scheme are memory overhead and computation overhead. This paper analyzes the ways to overcome these drawbacks. The modified Blom's scheme proposes the use of random matrix as the public matrix instead of the Vandermonde matrix. This reduces the computational complexity corresponding to the columns in Vandermonde matrix.

In this paper, we have performed an analysis between the original Blom's scheme and the proposed Blom's scheme with respect to computational effort. Different simulations have been made to a node size of 2,4,6, 8,10,20,30,40,50,100 and 200. Also, the Blom's scheme is made dynamic by changing the values of the secret matrix in order to protect the network from the malicious nodes. Different handshakes have been introduced so that the Central Authority can monitor the network at regular intervals.

The results of the simulation show that the computational complexity of the modified Blom's scheme is less when compared to that of the original Blom's scheme.

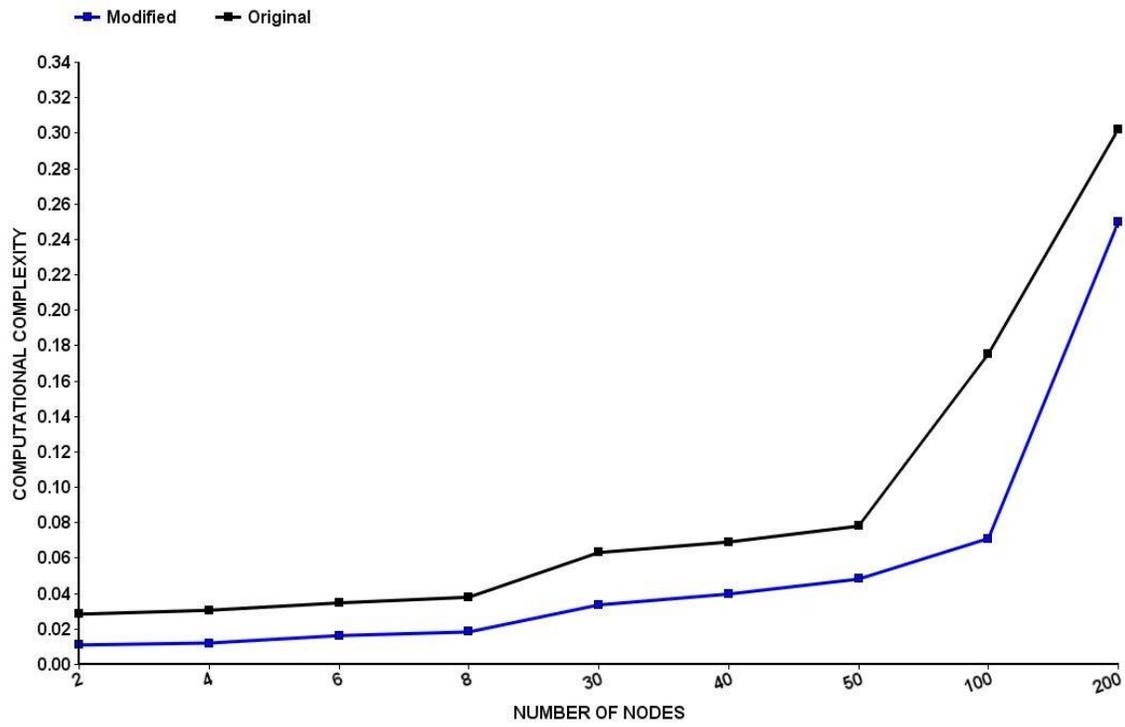

Figure-11: Comparison of Original scheme and modified scheme

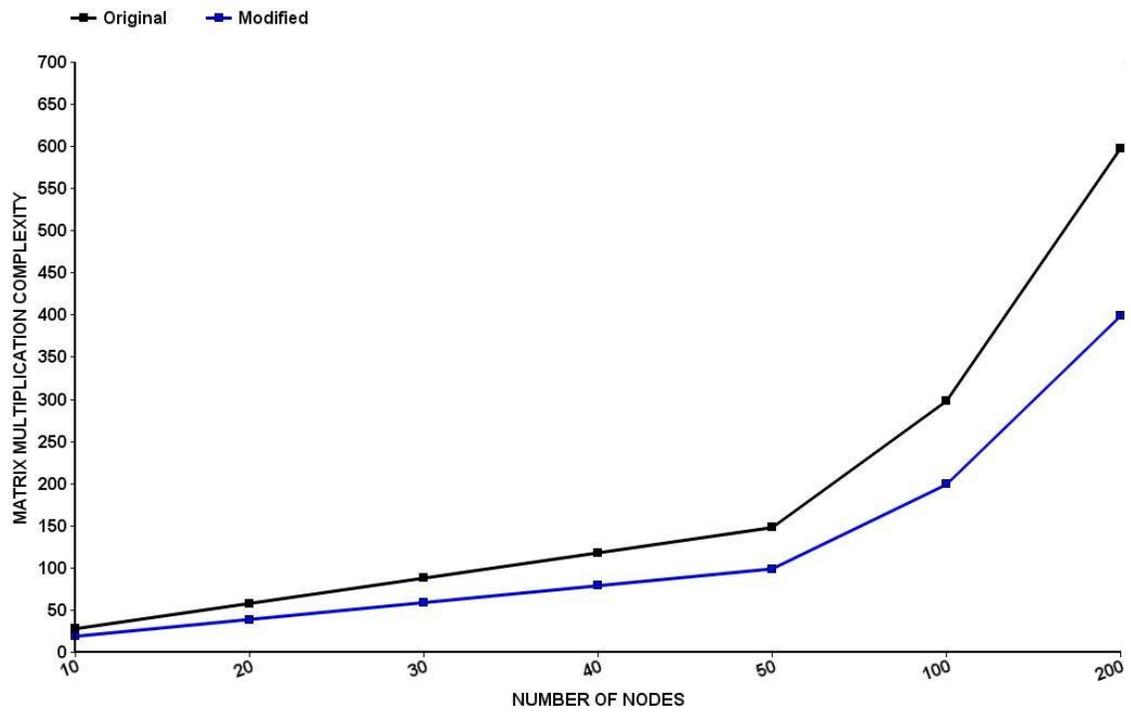

Figure-12: Matrix Multiplication Complexity

**CONCLUSION:**

The paper has shown that changes in the implementation of Blom's scheme provides new ways to protect the network from malicious nodes. An algorithm for dynamic changes of the keys is developed. Also, simulations are carried out for different node sizes and analyzed from the computational complexity point of view. We have shown how the protocols proposed for dynamic changing of keys can be run in an efficient manner.